\documentclass[prd,twocolumn,floatfix,preprintnumbers]{revtex4}

\usepackage{graphicx}
\usepackage{amsmath}
\usepackage{graphicx,epsfig}
\usepackage{bm}
\usepackage{latexsym,amssymb,amsmath,float}

\newcommand {\ga} {\ {\raise-.5ex\hbox{$\buildrel>\over\sim$}}\ }
\newcommand {\la} {\ {\raise-.5ex\hbox{$\buildrel<\over\sim$}}\ }

\def\be{\begin{equation}}
\def\ee{\end{equation}}
\def\ba{\begin{eqnarray}}
\def\ea{\end{eqnarray}}

\renewcommand{\[}{\left[}
\renewcommand{\]}{\right]}

\begin{document}

\title{A new generic evolution for $k$-essence dark energy with $w \approx -1$}
\author{John Kehayias}
\affiliation{Center for Programs in Contemporary Writing,
University of Pennsylvania, Philadelphia, PA  ~~19104}
\author{Robert J. Scherrer}
\affiliation{Department of Physics and Astronomy, Vanderbilt University,
Nashville, TN  ~~37235}

\begin{abstract}
We reexamine $k$-essence dark energy models with a scalar field $\phi$ and a factorized Lagrangian,
$\mathcal L = V(\phi)F(X)$, with
$X = \frac{1}{2} \nabla_\mu \phi \nabla^\mu \phi.$  A value of the equation of state parameter, $w$,
near $-1$ requires either $X \approx 0$ or $dF/dX \approx 0$.  Previous work showed
that thawing models with $X \approx 0$ evolve along a set of unique trajectories for $w(a)$, while
those with $dF/dX \approx 0$ can result in a variety of different forms for $w(a)$.  We
show that if $dV/d\phi$ is small and $(1/V)(dV/d\phi)$ is roughly constant, then the latter models also
converge toward a single unique set of behaviors for $w(a)$, different from those with $X \approx 0$.
We derive the functional form for $w(a)$ in this case, determine the conditions on $V(\phi)$ for which it applies,
and present observational constraints on this new class of models.  We note that $k$-essence
models with $dF/dX \approx 0$ correspond to a dark energy sound speed $c_s^2 \approx 0$.

\end{abstract}

\maketitle

\section{Introduction}

Observational evidence \cite{union08,hicken,Amanullah,Union2,Hinshaw,Ade,Betoule}
indicates that roughly
70\% of the energy density in the
universe is in the form of a component called dark energy, which has negative pressure,
and roughly 30\% is in the form of nonrelativistic matter.
The dark energy component can be parametrized in terms of its equation of state parameter, $w$,
defined as the ratio of the dark energy pressure to its density:
\be
\label{w}
w=p/\rho.
\ee
A cosmological constant, $\Lambda$, corresponds to the case $\rho = constant$ and $w = -1$ .

While a model with a cosmological constant and cold dark matter ($\Lambda$CDM) is consistent
with current observations,
there are other models of dark energy that have a dynamical equation
of state.
The most widely-investigated are
quintessence models, with a time-dependent scalar field, $\phi$,
having potential $V(\phi)$
\cite{RatraPeebles,Wetterich,Ferreira,CLW,CaldwellDaveSteinhardt,Liddle,SteinhardtWangZlatev}.
(See Ref. \cite{Copeland1} for a review).

While quintessence generically produces a time-varying value for
$w$, 
a successful model must closely mimic $\Lambda$CDM in order to be consistent
with current observations. Hence, a viable model should yield a present-day value
of $w$ close to $-1$.
This fact has been exploited in a number of papers that explored the evolution of a scalar field
subject to the constraint that $w$ must be close to $-1$
\cite{ScherrerSen,ds1,Chiba,ds2,ds3,Swaney}.  By imposing this constraint, one can reduce an
infinite number of models to a finite set of behaviors for $w(a)$.

In Ref. \cite{CBS}, this methodology was extended to $k$-essence models, which are characterized by a non-standard
kinetic term in the Lagrangian.  Ref. \cite{CBS} found two sets of solutions that yield $w \approx -1$.  The first corresponds to
$\dot \phi \rightarrow 0$ (where dot will refer throughout to the time derivative),
and it yields a single set of behaviors for $w(a)$.  The evolution of $w$ in this case turns out to
be identical to the quintessence models investigated in Refs. \cite{ds1,Chiba,ds2}.  The second solution corresponds to $\dot \phi \rightarrow
constant$.  However, in the latter case, the solution is sensitive to the functional form for $V(\phi)$ and therefore
fails to correspond to a single set of behaviors for $w(a)$.

In this paper, we revisit the second class of these solutions and show that, under some conditions on the potential $V(\phi)$, they do converge to a
single unique set of trajectories for $w(a)$.  Specifically, when $|(1/V)(dV/d\phi)|$ is small and nearly constant as
$\phi$ evolves, then the evolution of $w(a)$ converges toward a single
functional behavior.  Furthermore, unlike the solutions derived in
Ref. \cite{CBS}, the new class of solutions derived here correspond to behavior
for $w(a)$ that differs from previously-examined quintessence evolution.

In the next section, we briefly review previously-derived results for quintessence and $k$-essence evolution for $w$ near
$-1$.  In Sec. III, we present our new results for $k$-essence evolution, along with a discussion of the parameter ranges
over which these solutions are valid.  We discuss our results, including observational constraints, in Sec. IV.

\section{Previous results}

Before deriving our new results for $k$-essence, we need to present, for comparison, previously-derived results for
both quintessence and $k$-essence evolution.  We assume a flat universe with the Hubble parameter
given by
\begin{equation}
\label{H}
H = \left(\frac{\dot{a}}{a}\right) = \sqrt{\rho/3}.
\end{equation}
Here $a$ is the scale factor (with $a=1$ at the present), $\rho$ is the total density, and we work in units
for which $8 \pi G = 1$.  At late times, the contribution of photons and neutrinos to the expansion
can be neglected, so we take $\rho$ to include only matter (dark matter plus baryons) with
a density scaling as $a^{-3}$, and our unknown dark energy component, with a density which we assume
to be approximately (but not exactly) constant.

\subsection{Quintessence}

In this section, we will assume that the dark energy is provided by a minimally-coupled
scalar field, $\phi$, with equation of motion given by
\begin{equation}
\label{qevol}
\ddot{\phi}+ 3H\dot{\phi} + \frac{dV}{d\phi} =0.
\end{equation}
Equation (\ref{qevol}) indicates
that the field rolls downhill in the potential $V(\phi)$,
but its motion is damped by a term proportional to $H$.

The pressure and density of the
scalar field are given by
\begin{equation}
p = \frac{\dot \phi^2}{2} - V(\phi),
\end{equation}
and
\begin{equation}
\label{rhodense}
\rho = \frac{\dot \phi^2}{2} + V(\phi),
\end{equation}
respectively, and the equation of state parameter, $w$,
is given by equation (\ref{w}).

We will consider only ``thawing" models, for which the scalar field
is initially at rest ($\dot \phi = 0$, $w = -1$) and rolls downhill in the potential $V(\phi)$ so that $w$ increases
up to the present \cite{CL}.  Then Ref. \cite{ScherrerSen} considered potentials satisfying the inflationary slow-roll
conditions, namely
\begin{equation}
\label{slow1}
\left(\frac{V^\prime}{V}\right)^2 \ll 1,
\end{equation}
and
\begin{equation}
\label{slow2}
\frac{V^{\prime\prime}}{V} \ll 1,
\end{equation}
where the prime indicates throughout derivatives with respect to the scalar field, $\phi$.

Note, however, that the solutions derived here differ markedly from the inflationary slow-roll solutions.
In the latter case, $H$ in Eq. (\ref{qevol}) contains only the density of the scalar field itself,
and a solution can be derived by setting $\ddot \phi$ in Eq. (\ref{qevol}) equal to zero.  When both
the matter and scalar field energy densities are included in $H$, this solution is no longer valid, as discussed
in detail in Refs. \cite{Linder1,Linder2}.

When conditions (\ref{slow1}) and (\ref{slow2}) are imposed on the potential, along with the thawing
initial condition ($\dot \phi = 0$ at early times), it is possible to derive an approximate analytic
solution for $w(a)$ that is independent of $V(\phi)$.  This solution is \cite{ScherrerSen}
\begin{equation}
\label{wlinear}
1+w(a) = (1+w_0)\frac{\left[G(a) - (G(a)^2-1) \coth^{-1} G(a) \right]^2}
{\left[ G(1) - (G(1)^2-1) \coth^{-1} G(1)  \right]^2},
\end{equation}
where $w_0$ is the value of $w$ at the present.
The function $G(a)$ is
\begin{equation}
G(a) = \sqrt{1+(\Omega_{\phi 0}^{-1} - 1)a^{-3}},
\end{equation}
where $\Omega_{\phi 0}$ is the fraction of the total density at present contributed by the scalar field,
which we will take throughout to be $\Omega_{\phi 0} = 0.7$.  With these
definitions, $G(a) = 1/\sqrt{\Omega_\phi(a)}$ and $G(1) = 1/\sqrt{\Omega_{\phi 0}}$.
Here and throughout we will not give detailed derivations of previously-derived results but will instead
cite the original papers; in this case,
a detailed derivation of Eq. (\ref{wlinear})
is given in Ref. \cite{ScherrerSen}.
Note that we use different notation and express our results in a different functional form than some of the
earlier works cited here, both for the sake of increased simplicity and to avoid confusion with previously-adopted $k$-essence
notation.  The function given by Eq. (\ref{wlinear}) is displayed in Fig. 1 (green, long-dashed curve).
\begin{figure}[tbh]
\centerline{\epsfxsize=4truein\epsffile{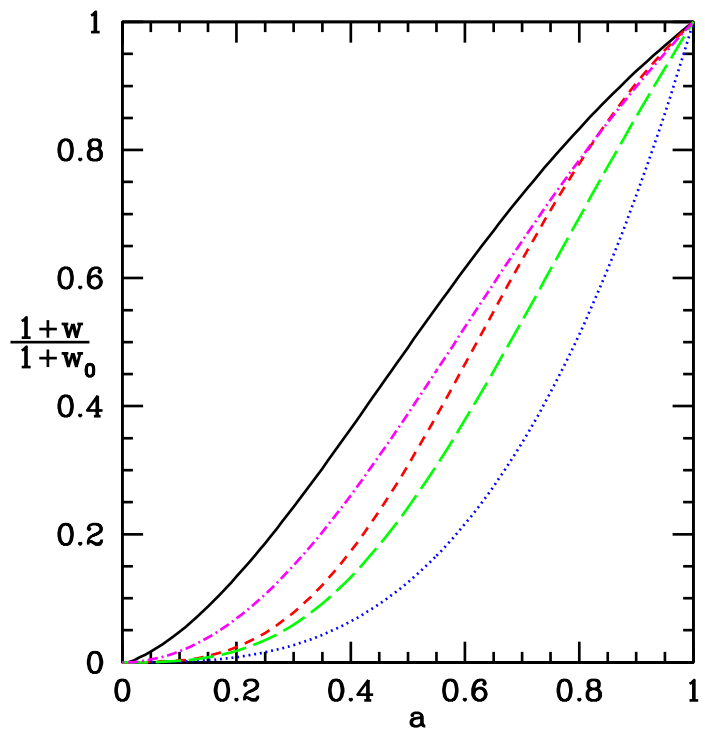}}
\caption{Evolution of $1+w$  relative to its value at the present, $1+w_0$,
as a function of the scale factor $a$
for the analytic predictions discussed in this paper.  Solid (black) curve is for $k$-essence
with $F_X \approx 0$ (the new result of this paper).  Blue (dotted) curve and red (short-dashed) curve are for $k$-essence
with $X \approx 0$ or quintessence with nonnegligible curvature in the potential, for $K=2$ and $K \rightarrow 0$,
respectively.  Green (long-dashed) curve is for quintessence in a nearly flat potential.
Magenta (dot-dashed) curve is for noncanonical quintessence with $\alpha = 2$.}
\end{figure}

In Refs. \cite{ds1,Chiba,ds2}, the condition on the potential given by Eq. (\ref{slow1}) was retained,
but condition (\ref{slow2}) was relaxed, resulting in a wider range of possible behaviors.  In this
case, the evolution of $w$ with scale factor is given by \cite{ds1,Chiba,ds2}
\begin{widetext}
\begin{equation}
\label{wquad}
1 + w(a) = (1+w_0)a^{3(K-1)}\frac{[(G(a)+1)^K(K-G(a))
+(G(a)-1)^K(K+G(a))]^2}
{[(G(1)+1)^K(K-G(1))
+(G(1)-1)^K (K+G(1))]^{2}},
\end{equation}
\end{widetext}
where the constant $K$ is a function of $V^{\prime \prime}/V$,
\begin{equation}
\label{Kdef}
K = \sqrt{1 - (4/3)V^{\prime \prime}(\phi_*)/V(\phi_*)},
\end{equation}
evaluated at $\phi_*$, which can be taken to be the initial value of $\phi$ \cite{Chiba}.
Now instead of a single functional form for $w(a)$ for a given value of $w_0$, Eq. (\ref{wquad})
provides a family of solutions that depend on $K$.  As $K$ becomes large, these solutions
thaw more slowly, i.e., $w$ remains close to $-1$ until later in the evolution \cite{ds1}.
In the opposite limit, as $K \rightarrow 1$, the solution in Eq. (\ref{wquad}) approaches
the evolution given in Eq. (\ref{wlinear}).  For $K \rightarrow 0$, $w$ increases more
rapidly than in Eq. (\ref{wlinear}). This behavior is illustrated in Fig. 1,
where $(1+w)/(1+w_0)$ is displayed as a function of $a$
for $K=2$ (blue, dotted curve) and $K \rightarrow 0$ (red, short-dashed curve).

\subsection{$k$-essence}

Now consider $k$-essence models with $w$ near $-1$.
In general, $k$-essence can be defined as
any scalar field $\phi$ with a noncanonical kinetic term,
so that the Lagrangian is of the form ${\mathcal L}(X,\phi)$,
where
\begin{equation}
\label{grad}
X = \frac{1}{2} \nabla_\mu \phi \nabla^\mu \phi.
\end{equation}
In practice, only a few special classes of such models have been explored
in detail.
The most widely-investigated class of models (and the
one examined in detail here and in Ref. \cite{CBS}) is taken 
to have a Lagrangian in factorized form:
\begin{equation}
\label{p}
{\mathcal L} = V(\phi)F(X).
\end{equation}
Such models were first introduced for inflation \cite{Arm1,Garriga}, and later extended to possible models for dark
energy \cite{Chiba1,Arm2,Arm3,Chiba2,Chimento1,Chimento2,Scherrer}.

Before considering such models in detail, we briefly mention a second class of models,
for which the Lagrangian has the form
\begin{equation}
\mathcal L = X^\alpha - V(\phi).
\end{equation}
These models have been dubbed ``noncanonical quintessence" and have been previously examined as
models both for inflation \cite{UST,RKK} and
for dark energy \cite{FLH,Unnikrishnan,Das,SahniSen,OGSS,LiScherrer}.
For these models, Li and Scherrer \cite{LiScherrer}
showed that when both slow-roll conditions on the potential (Eqs. \ref{slow1} and
\ref{slow2}) are satisfied, the equation of state is well-approximated by 
\begin{widetext}
\begin{equation}
\label{noncanon}
1+w(a) = (1+w_0)\frac{\left[G(a) - (G(a)^2-1) \coth^{-1} G(a) \right]^{2\alpha/(2\alpha-1)}}
{\left[ G(1) - (G(1)^2-1) \coth^{-1} G(1)  \right]^{2\alpha/(2\alpha-1)}}.
\end{equation}
\end{widetext}
As expected, this expression for $1+w(a)$ reduces to the corresponding quintessence result
(Eq. \ref{wlinear}) when $\alpha=1$, which corresponds to quintessence with a standard kinetic term.
The behavior of $(1+w)/(1+w_0)$ as a function of $a$ for the representative case $\alpha = 2$ is
shown in Fig. 1 (magenta, dot-dashed curve).

Now we direct our attention to factorizable $k$-essence models with the Lagrangian
given by Eq. (\ref{p}); such models are what is usually meant by the term ``$k$-essence."
The pressure in these models is simply given by equation (\ref{p}), while the energy density is
\begin{equation}
\label{rho}
\rho = V(\phi)[2X F_X - F],
\end{equation}
where $F_X \equiv dF/dX$.
Therefore, the equation of state parameter is
\begin{equation}
\label{wk}
w = \frac{F}{2X F_X - F}.
\end{equation}
The sound speed, which is relevant for the growth of density perturbations,
is
\begin{equation}
\label{cs}
c_s^2 =  \frac{F_X}{2XF_{XX} + F_X},
\end{equation}
with $F_{XX} \equiv d^2F/dx^2$.
In the flat Robertson-Walker metric, the equation for the evolution of the
$k$-essence field takes the
form:
\begin{equation}
\label{kevol}
(F_X + 2X F_{XX})\ddot \phi + 3H F_X \dot \phi + (2XF_X - F)\frac{V^\prime}{V} =
0.
\end{equation}

Chiba et al. \cite{CBS}
noted that $w \approx -1$ in Eq. (\ref{wk}) requires
\begin{equation}
\label{wcondition}
|XF_X| \ll |F|,
\end{equation}
which can be satisfied when either (i) $X \approx 0$ or (ii)
$F_X \approx 0$.  Note that these two conditions are sufficient, but not necessary to produce
$w \approx -1$; one can derive other functional forms for $F(X)$ for which Eq. (\ref{wcondition}) is satisfied for arbitrary
$X$.  For example, if $F = X^{-\alpha}$ we obtain
\begin{equation}
w  = -\frac{1}{2\alpha + 1},
\end{equation}
and $\alpha << 1$ corresponds to $w \approx -1$.  Here we will consider only the two cases examined
in Ref. \cite{CBS}, since these both converge toward unique sets of behaviors for $w(a)$.

Consider first case (i).  For this case, Chiba et al. showed
that the resulting evolution for $w$ is given by
\begin{widetext}
\begin{equation}
\label{wkX0}
1 + w(a) = (1+w_0)a^{3(K-1)}\frac{[(G(a)+1)^K(K-G(a))
+(G(a)-1)^K(K+G(a))]^2}
{[(G(1)+1)^K(K-G(1))
+(G(1)-1)^K (K+G(1))]^{2}},
\end{equation}
\end{widetext}
where now,
\begin{equation}
K = \sqrt{1 - \frac{4}{3}\frac{V^{\prime \prime}(\phi_i)}{F_X(0)V(\phi_i)^2}}.
\end{equation}
This result is identical to the corresponding quintessence result
in Eq. (\ref{wquad}).  Thus, these two models are observationally indistinguishable.
The behavior of $(1+w)/(1+w_0)$ as given by Eq. (\ref{wkX0}) is displayed in Fig. 1
for $K=2$ (blue, dotted curve) and $K \rightarrow 0$ (red, short-dashed curve).

For case (ii), Chiba
et al. derived a functional form for $w(a)$, but the result depends on $V(\phi)$ and is therefore considerably
less interesting.  It is this second case that we will revisit in the next section, showing that there are some conditions
under which it produces a single functional behavior for $w(a)$ that is independent of $V(\phi)$.

\section{Evolution of $w$ for $k$-essence models with $F_X \approx 0$}

Consider a $k$-essence model for which $F_X \approx 0$.  Following
Ref. \cite{CBS}, we will expand $F(X)$ around the extremum in $F$, which we
will take to occur at $X = X_m$.  Then taking
\begin{equation}
\label{deltadef}
X = X_m + \Delta,
\end{equation}
where $\Delta \ll X_m$, we can
write $F(X)$ as
\begin{equation}
\label{Fexp}
F(X)=F(X_m)+\frac12 F_{XX}(X_m)\Delta^2,
\end{equation}
so that
\begin{eqnarray}
\label{FXexp}
F_X(X) &=& F_{XX}(X_m) \Delta,\\
\label{FXXexp}
F_{XX}(X) &=& F_{XX}(X_m).
\end{eqnarray}
Then Eq. (\ref{wk}) can be expanded to linear order in $\Delta$ to yield
\begin{equation}
\label{wexpand}
1+w = \[\frac{2 X_m
F_{XX}(X_m)}{F(X_m)}\]\Delta.
\end{equation}
In order to solve for $w(a)$,  we first need to reexpress Eq. (\ref{kevol}) in terms of $\Delta$
instead of $\phi$. Using Eqs. (\ref{deltadef}) - (\ref{FXXexp}), we
can rewrite Eq. (\ref{kevol}), up to linear order in $\Delta$, as
\begin{widetext}
\begin{equation}
\label{Deltaeq1}
\dot \Delta + 3H \Delta +
\left(\sqrt{2 X_m}\frac{V^\prime}{V}\right)\Delta
-\left(\sqrt{2 X_m}\frac{V^\prime}{V}\right)\left( \frac{F(X_m)}{4 X_m^2 F_{XX}(X_m)}\right)\Delta
- \left(\sqrt{2X_m} \frac{V^\prime}{V}\right)\left(\frac{F(X_m)}{2 X_m F_{XX}(X_m)}\right) = 0.
\end{equation}
\end{widetext}
The ratio of the third term to the final term is (from Eq. \ref{wexpand}) equal to $1+w$,
which we take to be $\ll 1$.  The ratio of the fourth term to the final term is $\Delta/2X_m$, and
we have assumed that $\Delta/X_m \ll 1$.  Thus, the third and fourth terms in Eq. (\ref{Deltaeq1}) are negligible compared to the final term
in that equation.  Then Eq. (\ref{Deltaeq1}) simplifies to
\begin{equation}
\label{Deltaeq2}
\dot \Delta + 3H \Delta
- \left(\sqrt{2X_m} \frac{V^\prime}{V}\right)\left(\frac{F(X_m)}{2 X_m F_{XX}(X_m)}\right) = 0.
\end{equation}
To solve this equation, we make one final assumption:  that $V^\prime/V$ is roughly constant as the $k$-essence
field evolves through the period of interest.  With this assumption,
Eq. (\ref{Deltaeq2}) can be solved exactly to yield
\begin{equation}
\label{Delta}
\Delta = \frac{C}{\sqrt{3 \rho_{\phi 0}}}[G(a) - [G(a)^2-1]\coth^{-1}G(a)],
\end{equation}
where $C$ is the negative of the third term in Eq. (\ref{Deltaeq2}), now taken to be constant:
\begin{equation}
\label{C}
C = \left(\sqrt{2X_m} \frac{V^\prime}{V}\right)\left(\frac{F(X_m)}{2 X_m F_{XX}(X_m)}\right).
\end{equation}
Then Eq. (\ref{wexpand}) gives the value of $1+w$:
\begin{equation}
\label{gammak}
1+w = \sqrt{\frac{2 X_m}{3 \rho_{\phi 0}}} \frac{V^\prime}{V}[G(a) - (G(a)^2-1)\coth^{-1}G(a)].
\end{equation}
We can reexpress this in terms of the $w_0$ as in Eqs. (\ref{wlinear}), (\ref{wquad}), (\ref{noncanon}), and (\ref{wkX0})
to give:
\begin{equation}
\label{wkF0}
1+w(a) = (1+w_0)\frac{G(a) - (G(a)^2-1) \coth^{-1} G(a)}
{G(1) - (G(1)^2-1) \coth^{-1} G(1)  }.
\end{equation}
Eq. (\ref{wkF0}) is the main result of this paper.

In Fig. 1, we show the behavior of $w(a)$ given by Eq. (\ref{wkF0}) (solid black curve), along with
the corresponding behavior for the models examined previously. Note that, unlike the solution for $k$-essence with $X \approx 0$,
the result here does not resemble any corresponding quintessence model, although it does correspond to
the limiting behavior of noncanonical quintessence (Eq. \ref{noncanon}) in the limit where
$\alpha \rightarrow \infty$.  This correspondence is not surprising, as $\alpha \rightarrow \infty$ in
noncanonical quintessence corresponds to the limit $X \rightarrow constant$ \cite{SahniSen}, the same behavior
as in the $k$-essence models considered here.

The difference between this result for $k$-essence and the corresponding behavior for quintessence (Eq. \ref{wlinear})
is particularly clear if we examine these results in the $w - w^\prime$ plane \cite{CL,wScherrer},
where $w^\prime \equiv a({dw}/{da})$,
in
the limit $a \ll 1$.  In that limit, Eq. (\ref{wlinear}) reduces
to $w^\prime = 3(1+w)$ for
quintessence (see also Ref. \cite{Linder2}), while Eq. (\ref{wkF0}) gives
$w^\prime = \frac{3}{2}(1+w)$ for $k$-essence.

Now consider the conditions on the model parameters necessary for Eq. (\ref{wkF0}) to represent a good approximation
to the evolution of $w$.
The conditions we imposed to derive Eq. (\ref{wkF0}) are: (i) $1+w \ll 1$, (ii)  $\Delta \ll X_m$, and (iii)  $V^\prime/V$ is approximately
constant as $\phi$ evolves.

Clearly, if all of the other parameters in the $k$-essence models are $\sim \mathcal{O}(1)$, then conditions (i) and (ii)
can be satisfied by choosing $(V^\prime/V)^2 \ll 1$ as in Eq. (\ref{slow1}); this follows directly from Eqs. (\ref{Delta})-(\ref{gammak}).
Condition (iii) indicates that $V^\prime/V$ evolves only a small amount compared to its initial value as $\phi$ evolves.  This
will be the case as long as $(V^\prime/V)^\prime/(V^\prime/V) \delta \phi \ll 1$, where $\delta \phi$ is the total change in
$\phi$ between $a=0$ and $a=1$.

In Fig. 2, we compare the analytic approximation of Eq. (\ref{wkF0}) to a numerical integration of the equation
for $k$-essence evolution, where the parameters of these models are chosen to satisfy $(V^\prime/V)^2 \ll 1$ and
$(V^\prime/V)^\prime/(V^\prime/V) \delta \phi \ll 1$; these conditions can be satisfied for all three potentials
by taking the initial value of $\phi$ to be sufficiently large.  For all of these cases we take $F(X) = F_0 + F_2 (X-X_m)^2$.
The fit to our
analytic expression is very good in all three cases, and nearly exact for the exponential potential.  The latter
is not surprising, as the exponential potential has $V^\prime/V = constant$ by construction.
\begin{figure}[tbh]
\centerline{\epsfxsize=4truein\epsffile{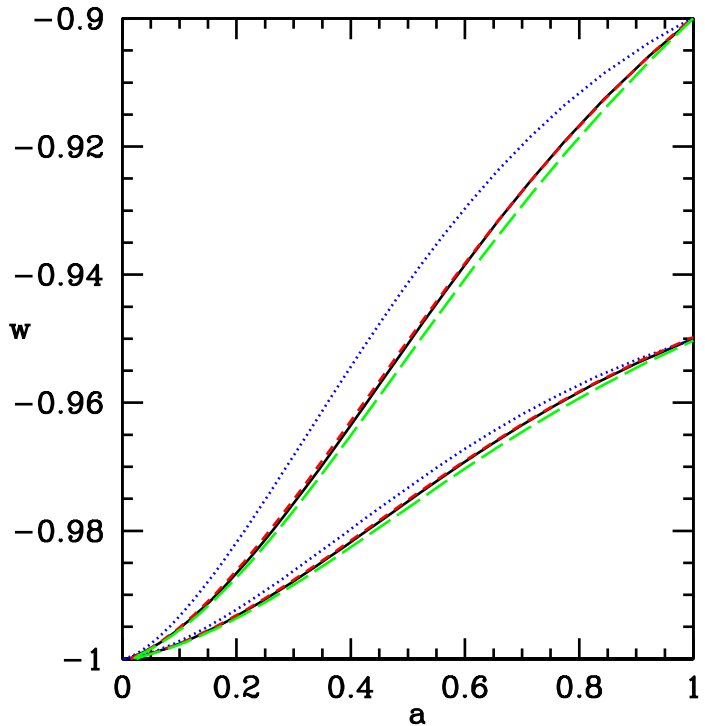}}
\caption{Evolution of $w$ as a function of $a$, normalized to $a=1$ at the present, with $\Omega_{\phi 0} = 0.7$
and $w_0 = -0.9$ and $-0.95$, for models with $F_X \approx 0$.  Solid (black) curve is
our analytic approximation (Eq. \ref{wkF0}).  Dotted (blue) curve is for $V(\phi) = V_0/\phi$,
short-dashed (red) curve is $V(\phi) = e^{-\lambda \phi}$, and long-dashed (green) curve
is $V(\phi) =  e^{-\phi^2/\sigma^2}$.  We take $F(X) = F_0 + F_2 (X-X_m)^2$ for all three cases.}
\end{figure}

\section{Discussion}

Now we can compare the behavior of $k$-essence models with $F_X \approx 0$ to those of Ref. \cite{CBS} with $X \approx 0$.
The $F_X \approx 0$ models yield a new form for the evolution of $w(a)$, distinct
from previous behaviors that have been derived for other models, while $X \approx 0$ models
correspond to behavior that is identical
to the results for quintessence evolution given in Refs. \cite{ds1,Chiba}.  On the other hand, our results
for $F_X \approx 0$ models
are applicable to a much more restricted set of scalar field potentials than is the case for
$X \approx 0$, namely, our results apply only to potentials for which
$(V^\prime/V)^\prime/(V^\prime/V) \delta \phi \ll 1$.
This is the reason that Chiba et al. \cite{CBS} found
a variety of possible behaviors for $w(a)$ with $F_X \approx 0$ (see Fig. 4 of Ref. \cite{CBS}); the potentials examined in that paper did
not satisfy our (very restrictive) conditions on $V(\phi)$.

Now consider the observational constraints on our model.  We will compare with the recent results
of Alam et al. \cite{Alam}, derived from baryon acoustic oscillation measurements from the
Sloan Digital Sky Survey III, cosmic microwave background observations from Planck,
and Type Ia supernovae
data.  Alam et al. express their constraints on $w$ in terms of the
Chevallier-Linder-Polarski (CPL) parametrization
\cite{CP,Linder}:
\begin{equation}
\label{CPL}
w = w_0 +(1-a)w_a,
\end{equation}
where $w_a$ and $w_0$ are constants, with $w_0$ being the present-day value of $w$.
The most stringent bounds on $w_a$ and $w_0$ in Alam et al. correspond to a narrow ellipse in the $w_0$, $w_a$ plane.  In this two-parameter
model, neither $w_0$ nor $w_a$ is individually strongly constrained, but a linear combination
of the two is tightly bounded.  The reason for this characteristic narrow elliptical bounded region in $w_0 - w_a$ space
is the existence of a pivot redshift $z_p$,
at which the errors on $w$ are minimized \cite{Albrecht}.  In particular,
Alam et al. \cite{Alam} find a pivot redshift of $z_p = 0.37$, at which  $w(z_p) = -1.05 \pm 0.05$.  

We can exploit the fact that our model and the other models discussed in this paper are
well-fit by the CPL parametrization for $a_p < a < 1$, and each model gives a unique prediction
for $w(a_p)$ as a function of $w_0$.
Hence, much
stronger constraints can be placed on these models than on a generic dark energy model; in particular,
we can derive a tight upper bound on the present-day value of $w$.  First note that
the pivot redshift $z_p$ is related to $a_p$ through $a_p = 1/(1+z_p)$,
so we have $a_p = 0.73$.  Then our $k$-essence model with $F_X \approx 0$ must satisfy the $2-\sigma$ upper bound $w(a = 0.73) < -0.95$.
We can then simply read off the allowed value of $w_0$ from Fig. 1, namely, $w_0 < -0.93$.

It is clear
from this argument that the models that allow the largest values of $w_0$ are those for which $w$
increases most rapidly from $a = 0.73$ to $a=1$.  Hence, our new $k$-essence model with $F_X \approx 0$
is the most strongly constrained of those displayed in Fig. 1.  In comparison, the quintessence
model with a nearly-flat potential \cite{ScherrerSen} yields the constraint $w_0 < -0.91$, while the least strongly-constrained
model is the $k$-essence model with $X \approx 0$ (or equivalently, the quintessence model
with non-negligible curvature in the potential) with $K = 2$, for which $w_0 < -0.87$.  Larger
values of K are even less strongly constrained \cite{ds1}.  (See Ref.
\cite{Linder3} for another
approach to observational constraints on thawing models).

Note further that the $k$-essence models considered here with $F_X \approx 0$
make a very different prediction for the dark energy sound speed than do the
previously-examined models with $X \approx 0$.  From Eq. (\ref{cs}),
we see that our models give $c_s^2 \approx 0$, while the $X \approx 0$ models
have $c_s^2 \approx 1$.  Current observations are unable to significantly
constrain $c_s$ for dark energy (see, e.g., Refs.
\cite{Weller,Bean,Hannestad,Heneka}), so these two extreme cases
are not currently distinguishable, but future experiments such as
Euclid \cite{Amendola} may provide useful constraints on the sound speed
of dark energy. 

In summary, we have derived a new generic thawing evolution of $k$-essence with $w$ near $-1$; this is essentially
a special case of the $F_X \approx 0$ solutions previously derived in Ref. \cite{CBS},
but for which additional constraints on the potential $V(\phi)$ yield a single
set of evolutionary behaviors for $w(a)$.  It is interesting that
$w$ in this model evolves away from $-1$ more rapidly than in any of the other models considered here, which allows
us to place tighter constraints on this model than on any of the others.  In constrast, the $k$-essence
models with $X \approx 0$
examined in Ref. \cite{CBS} require fewer conditions on the potential $V(\phi)$ and are less tightly constrained
by observations.  Our $F_X \approx 0$ models also provide a simple case for which $w \approx -1$, but
the dark energy sound speed is close to zero.

\section*{Acknowledgments}
R.J.S. was supported in part by the Department of Energy
(DE-SC0019207).


\begin{thebibliography}{99}

\bibitem{union08}
  M.~Kowalski {\it et al.},
  \apj  {\bf 686}, 749 (2008).
  
\bibitem{hicken}
M.~Hicken {\it et al.},
\apj {\bf 700}, 1097 (2009).

\bibitem{Amanullah}
R. Amanullah {\it et al.},
\apj {\bf 716}, 712 (2010).

\bibitem{Union2}
N. Suzuki {\it et al.},
\apj {\bf 746}, 85 (2012).

\bibitem{Hinshaw}
G. Hinshaw, {\it et al.}, \apj Suppl. {\bf 208}, 19 (2013).
  
\bibitem{Ade}
P.A.R. Ade, {\it et al.},
Astron. Astrophys. {\bf 571}, A16 (2014).
 
\bibitem{Betoule}
M. Betoule {\it et al.},
Astron. Astrophys. {\bf 568}, A22 (2014).

   
\bibitem{RatraPeebles}
  B. Ratra and P.J.E. Peebles,
  Phys.\ Rev.\  D {\bf 37}, 3406 (1988).
  
\bibitem{Wetterich}
C. Wetterich, Astron. Astrophys. {\bf 301}, 321 (1995).
  
\bibitem{Ferreira} P.G. Ferreira and M. Joyce,
\prl {\bf 79}, 4740 (1997).

\bibitem{CLW} E.J. Copeland, A.R. Liddle, and D. Wands,
\prd{\bf 57}, 4686 (1998).  
  

\bibitem{CaldwellDaveSteinhardt}
  R.R. Caldwell, R. Dave and P. J. Steinhardt,
  Phys.\ Rev.\ Lett.\  {\bf 80}, 1582 (1998).
  

\bibitem{Liddle}
  A.R. Liddle and R.J. Scherrer,
  Phys.\ Rev.\  D {\bf 59}, 023509 (1999).
  
  
\bibitem{SteinhardtWangZlatev}
  P.J. Steinhardt, L.M. Wang and I. Zlatev,
  Phys.\ Rev.\  D {\bf 59}, 123504 (1999).
  
\bibitem{Copeland1}
E.J. Copeland, M. Sami, and S. Tsujikawa, Int. J. Mod. Phys. D
{\bf 15}, 1753 (2006).

\bibitem{ScherrerSen}
R.~J.~Scherrer and A.~A.~Sen,
Phys.\ Rev.\  D {\bf 77}, 083515 (2008)
 
\bibitem{ds1}
S.~Dutta and R.~J.~Scherrer,
\prd {\bf 78}, 123525 (2008).

\bibitem{Chiba}
T. Chiba, \prd {\bf 79}, 083517 (2009).
  

\bibitem{ds2}
  S.~Dutta, E.~N.~Saridakis and R.~J.~Scherrer,
  \prd {\bf 79}, 103005 (2009).

\bibitem{ds3}
S. Dutta and R.J. Scherrer,
Phys. Lett. B {\bf 704}, 265 (2011).

\bibitem{Swaney}
J.R. Swaney and R.J. Scherrer,
\prd {\bf 91}, 123525 (2015).

\bibitem{CBS}
T. Chiba, S. Dutta, and R.J. Scherrer,
\prd {\bf 80}, 043517 (2009).

\bibitem{CL}
R.R. Caldwell and E.V. Linder, \prl {\bf 95}, 141301 (2005).

\bibitem{Linder1}
E.V. Linder, \prd {\bf 73}, 063010 (2006).

\bibitem{Linder2}
R.N. Cahn, R. de Putter, and E.V. Linder,
JCAP {\bf 11}, 015 (2008).

\bibitem{Arm1}
C. Armendariz-Picon, T. Damour, and V. Mukhanov, Phys. Lett. B {\bf 458}, 209 (1999).

\bibitem{Garriga}
J. Garriga and V.F. Mukhanov,
Phys. Lett. B {\bf 458}, 219 (1999).

\bibitem{Chiba1}
T. Chiba, T. Okabe, M. Yamaguchi,
\prd {\bf 62}, 023511 (2000).

\bibitem{Arm2}
C. Armendariz-Picon, V. Mukhanov, and P.J. Steinhardt,
\prl {\bf 85}, 4438 (2000).

\bibitem{Arm3}
C. Armendariz-Picon, V. Mukhanov, and P.J. Steinhardt,
\prd {\bf 63}, 103510 (2001).

\bibitem{Chiba2}
T. Chiba, \prd {\bf 66}, 063514 (2002).

\bibitem{Chimento1}
L.P. Chimento and A. Feinstein, Mod. Phys. Lett. A {\bf 19}, 761 (2004).

\bibitem{Chimento2}
L.P. Chimento, \prd {\bf 69}, 123517 (2004).

\bibitem{Scherrer}
R.J. Scherrer, \prl {\bf 93}, 011301 (2004).

\bibitem{UST} S. Unnikrishnan, V. Sahni, and A. Toporensky, JCAP {\bf 8}, 018 (2012).
\bibitem {RKK} K. Rezazadeh, K. Karami, and P. Karani, JCAP {\bf 9}, 053 (2015).

\bibitem{FLH} W. Fang, H.Q. Lu, and Z.G. Huang, Class. Quant. Grav. {\bf 24}, 3799 (2007).
\bibitem{Unnikrishnan} Unnikrishnan, Phys.Rev.D {\bf78}, 063007 (2008).
\bibitem{Das} S. Das and A. Al Mamon, Astrophys. Space Sci. {\bf 355}, 371 (2015).
\bibitem{SahniSen} V. Sahni and A.A. Sen, Eur. Phys. Jour. C {\bf 77}, 225 (2017).
\bibitem{OGSS} Z. Ossoulian, T. Golanbari, H. Skeikhahmadi, and Kh. Saaidi, Adv. High Energy Phys. 3047461 (2016).
\bibitem{LiScherrer} D. Li and R.J. Scherrer, \prd {\bf 93}, 083509 (2016).

\bibitem{wScherrer} R.J. Scherrer, \prd {\bf 73}, 043502 (2006).

\bibitem{Alam}
S. Alam, et al., MNRAS {\bf 470}, 2617 (2017).

\bibitem{CP}
M. Chevallier and D. Polarski,
Int. J. Mod. Phys. D {\bf 10},
213 (2001).

\bibitem{Linder}
E.V. Linder, \prl {\bf 90}, 091301 (2003).

\bibitem{Albrecht}
A. Albrecht, et al., astro-ph/0609591.

\bibitem{Linder3}
E.V. Linder, \prd {\bf 91}, 063006 (2015).

\bibitem{Weller}
J. Weller and A.M. Lewis, MNRAS {\bf 346}, 987 (2003).

\bibitem{Bean}
R. Bean and O. Dore, \prd {\bf 69}, 083503 (2004).

\bibitem{Hannestad}
S. Hannestad, \prd {\bf 71}, 103519 (2005).


\bibitem{Heneka}
C. Heneka, et al., MNRAS {\bf 473}, 3882 (2018).


\bibitem{Amendola}
L. Amendola, et al., Living Rev. Rel. {\bf 16}, 6 (2013).

\end{thebibliography}
\end{document}